\documentclass[english,american]{article}
\usepackage[T1]{fontenc}
\usepackage[utf8]{inputenc}
\usepackage{parskip}
\usepackage{geometry}
\usepackage{setspace}
\makeatletter

\providecommand{\tabularnewline}{\\}

\makeatother

\usepackage{babel}
\begin{document}
\title{A Modified Satterthwaite (1941,1946) Effective Degrees of Freedom
Approximation }
\author{Matthias von Davier (vondavim@bc.edu)}
\date{November 22nd, 2024}
\maketitle
\begin{abstract}
This study introduces a correction to the approximation of effective
degrees of freedom as proposed by Satterthwaite (1941, 1946), specifically
addressing scenarios where component degrees of freedom are small.
The correction is grounded in analytical results concerning the moments
of standard normal random variables. This modification is applicable
to complex variance estimates that involve both small and large degrees
of freedom, offering an enhanced approximation of the higher moments
required by Satterthwaite's framework. Additionally, this correction
extends and partially validates the empirically derived adjustment
by Johnson \& Rust (1992), as it is based on theoretical foundations
rather than simulations used to derive empirical transformation constants.
\end{abstract}

\section{Introduction}

This article presents a correction to the approximation of effective
degrees of freedom as proposed by Satterthwaite (1941, 1946), specifically
for cases where the component degrees of freedom are small. Satterthwaite's
original work introduced an equation for estimating effective degrees
of freedom in scenarios involving complex variance estimators, which
rely on weighted sums of mean squared errors. 

The correction developed here is grounded in analytical results related
to the moments of standard normal random variables. In instances where
the components of complex variance estimators exhibit small degrees
of freedom, this correction offers a more accurate approximation of
the higher moments required by Satterthwaite's methodology. Furthermore,
this correction extends and partially justifies the empirically derived
adjustment proposed by Johnson \& Rust (1992), as it is based on theoretical
results rather than simulations used to derive empirical transformation
constants. 

While Satterthwaite's approach is widely used, it is recognized that
it relies on a number of assumptions. One of these is that the components
of the complex variance estimator are (approximately) independent
and that the degrees of degrees of freedom of each component are at
least moderate. If that is not the case, Satterthwaite (1946) noted
that the approximation may be poor. The case of dependent variance
components was tackled by Nedelman \& Jia (1998). Here we address
the common case that each of the variance components in a complex
estimate of variance has small or even only single degrees of freedom. 

The primary objective of this work is to provide a formula for effective
degrees of freedom that is applicable beyond the NAEP context, which
underpinned Johnson \& Rust's simulations. This more general adjustment
enables the estimation of effective degrees of freedom in various
applications, including analyses of data from surveys and assessment
programs that utilize variance estimates based on resampling methods.

\section{Sums of Independent Components in Complex Estimates of Variance}

Satterthwaite (1941, 1946) discusses complex variance estimates. For
this article, we consider the following situation: For $k=1,...,K,$and
$i=1,..,n_{k}$ denote $X_{ik}\sim N\left(\mu,\sigma^{2}\right)$
i.i.d. random variables. Then let $M_{k}=\frac{1}{n_{k}}\sum_{i=1}^{n_{k}}x_{ik}$
denote the mean for sample $k$, which implies $E\left(M_{k}\right)=\mu$,
and let
\[
S_{k}^{2}=\frac{n_{k}}{n_{k}-1}\sum_{i=1}^{n_{k}}\frac{\left(x_{ik}-M_{k}\right)^{2}}{n_{k}}
\]
denote the estimate of variance for sample $k$, with $E\left(S_{k}^{2}\right)=\sigma^{2}$
for all $k$. Let $\nu_{k}=n_{k}-1$ denote the degrees of freedom
of the variance estimate $S_{k}^{2}$. Then we have 
\[
X_{k}^{2}=\nu_{k}\frac{S_{k}^{2}}{\sigma^{2}}\sim\chi_{n_{k}-1}^{2}
\]
and hence $E\left(X_{k}^{2}\right)=\nu_{k}$ and $Var\left(X_{k}^{2}\right)=2\nu_{k}.$

For the complex estimate of variance, we consider the sum of the $S_{k}^{2}$,
\[
S_{*}^{2}=\sum_{k=1}^{K}S_{k}^{2}.
\]

Then we have for the expectation 
\[
E\left(S_{*}^{2}\right)=E\left[\sum_{k=1}^{K}S_{k}^{2}\right]=\sum_{k=1}^{K}E\left(S_{k}^{2}\right)=K\sigma^{2}.
\]

If the $S_{k}^{2}$ are independent, we can write
\[
Var\left(S_{*}^{2}\right)=Var\left[\sum_{k=1}^{K}S_{k}^{2}\right]=\sum_{k=1}^{K}Var\left(S_{k}^{2}\right)
\]

\section{Two Useful Identities}

\subsection{First Identity:}

Note that for any chi-square distributed variance estimate 
\[
S^{2}=\frac{\nu+1}{\nu}\sum_{i=1}^{\nu+1}\frac{\left(X_{i}-\overline{X}_{*}\right)^{2}}{\nu+1}
\]
with variance $\sigma_{*}^{2}$ we have for the variance of the chi-squared
\[
Var\left[\nu\frac{S^{2}}{\sigma^{2}}\right]=2\nu
\]
so that 
\[
\frac{\nu^{2}}{\sigma^{4}}Var\left(S^{2}\right)=2\nu\leftrightarrow\frac{Var\left(S^{2}\right)}{\sigma^{4}}=\frac{2}{\nu}
\]
which yields
\[
Var\left(S^{2}\right)=\frac{2\sigma^{4}}{\nu}
\]

\subsection{Expectation of a Squared Sum of Squared Standard Normal Random Variables}

Recall that for chi-squared random variables $\chi_{v}^{2}=\sum_{i=1}^{\nu}Z_{i}^{2}$
we have
\[
E(\chi_{v}^{2})=\nu,Var(\chi_{v}^{2})=2\nu.
\]

Plugging in the well-known identity $Var\left(X\right)=E\left(X^{2}\right)-\left[E\left(X\right)\right]^{2}$,
we find
\[
2\nu=Var\left(\left[\sum_{k=1}^{\nu}Z_{k}^{2}\right]\right)=E\left(\left[\sum_{k=1}^{\nu}Z_{k}^{2}\right]^{2}\right)-\left[E\left(\sum_{k=1}^{\nu}Z_{k}^{2}\right)\right]^{2}
\]
so that 
\[
E\left(\left[\sum_{k=1}^{\nu}Z_{k}^{2}\right]^{2}\right)=Var(\chi_{v}^{2})+\left[E(\chi_{v}^{2})\right]^{2}=2\nu+\nu^{2}=\left(\nu+2\right)\nu,
\]
so that for random variables $S_{k}=\sigma Z_{k}$ with $Z_{k}\sim N\left(0,1\right)$
and $\sigma>0$ we have
\[
E\left(\left[\sum_{k=1}^{K}S_{k}^{2}\right]^{2}\right)=\sigma^{4}\nu\left(\nu+2\right).
\]

\section{Main Idea of the Satterthwaite Approach}

It does not follow automatically that $S_{*}^{2}$ is chi-squared
if it is defined as a sum of mean squared differences. However, it
is a useful approach to assume the distribution of $S_{*}^{2}$ can
be approximated by a chi-square $\chi_{\nu_{?}}^{2}$distribution
with unknown degrees of freedom $\nu_{?}$. To obtain an estimate
of this quantity, the idea is to look a the 'useful identity' introduced
above, and to use the result 

\[
Var\left(S_{*}^{2}\right)=\frac{2\sigma_{*}^{4}}{\nu_{?}}\leftrightarrow\frac{2\left(\sigma_{*}^{2}\right)^{2}}{Var\left(S_{*}^{2}\right)}=\nu_{?}
\]
in order to estimate or approximate the unknown degrees of freedom
$\nu_{?}.$ For the sake of estimating $\nu_{?},$Satterthwaite (1946)
assumes that the $K$ components used estimate the variance $S_{k}^{2}$
are independent. Then, for this independent sum, the 'useful result'
is applied to obtain
\[
Var\left(S_{*}^{2}\right)=\sum_{k=1}^{K}Var\left(S_{k}^{2}\right)=\sum_{k=1}^{K}\frac{2\left(\sigma_{k}^{2}\right)^{2}}{\nu_{k}}.
\]

\section{An Estimate of the Effective Degrees of Freedom $\nu_{?}$}

The above result can then be applied to obtain
\[
\nu_{?}=\frac{2\left(\sigma_{*}^{2}\right)^{2}}{\sum_{k=1}^{K}\frac{2\left(\sigma_{k}^{2}\right)^{2}}{\nu_{k}}}=\frac{\left(\sigma_{*}^{2}\right)^{2}}{\sum_{k=1}^{K}\frac{\left(\sigma_{k}^{2}\right)^{2}}{\nu_{k}}}
\]

The main idea is to replace the true variance by an estimate of that
variance, namely, to approximate
\[
\left(\sigma_{Q}^{2}\right)^{2}\approx\left(S_{Q}^{2}\right)^{2}
\]
for both cases $Q=k$ and $Q=*$. The first step is replacing 
\[
\left(\sigma_{*}^{2}\right)^{2}\approx\left(S_{*}^{2}\right)^{2}
\]
and then 
\[
\sum_{k=1}^{K}\frac{\left(\sigma_{k}^{2}\right)^{2}}{\nu_{k}}\approx\sum_{k=1}^{K}\frac{\left(S_{k}^{2}\right)^{2}}{\nu_{k}}
\]
This plugging in of the estimates produces the Satterthwaite (1946)
equation
\[
\nu_{?}\approx\frac{\left(\sum_{k=1}^{K}S_{k}^{2}\right)^{2}}{\sum_{k=1}^{K}\frac{\left(S_{k}^{2}\right)^{2}}{\nu_{k}}}.
\]

\section{Some Properties of the Approximation}

Assume $S_{k}^{2}=S_{j}^{2}=C$ for all $k,j\in\left\{ 1,...,K\right\} $.
then we have 
\[
\frac{1}{\nu_{?}}=\frac{C^{2}\sum_{k}\frac{1}{\nu_{k}}}{K^{2}C^{2}}=\frac{1}{K^{2}}\sum_{k=1}^{K}\frac{1}{\nu_{k}}
\]
 or
\[
\frac{K^{2}}{\nu_{?}}=\sum_{k=1}^{K}\frac{1}{\nu_{k}}
\]
 Assume $\nu_{k}=\nu_{j}=\nu$. Then we have 
\[
\nu_{?}\approx\frac{K^{2}C^{2}}{C^{2}\sum_{k}\frac{1}{\nu_{k}}}=\frac{K^{2}}{K\frac{1}{\nu}}=K\nu
\]

With special case $\nu=1$ and all $S_{k}^{2}=S_{j}^{2}=C$ then $\nu_{?}=K$.
If $S_{j}^{2}=C$ and $S_{k}^{2}=0$ for $k\ne j$ we find 
\[
\nu_{?}=\frac{C^{2}}{\frac{C^{2}}{\nu_{j}}}=\nu_{j}
\]
and if $\nu_{j}=1$ we have $\nu_{?}=1$ in this case.

Hence, we can conclude that if all $\nu_{k}=1$ for $k=1,...,K$ we
have 
\[
1\le\nu_{?}\le K
\]
 since the function is smooth in the $S_{k}^{2}$. The maximum is
attained if all $S_{k}^{2}$are the same.

\section{Johnson \& Rust Correction for Jackknife Based Estimates}

$Satterthwaite$ (1941, 1946) mentioned that the approximation is
best applied when the $\nu_{k}$ are large, and that for small $\nu_{k},$the
approximation may not be as stable. Johnson \& Rust (1992) developed
an adjustment to overcome this limitation, based on a simulation and
empirically derived constants for the NAEP assessment program\footnote{It is important to note that the author has access to an unpublished
draft from the second author (Rust) as the proceedings submission
cited as Johnson \& Rust (1992) was apparently never completed and
is unavailable.}. The adjustment proposed in this work is used in a slightly simplified
form, until today, in the NAEP analyses. The adjustment formula in
the unpublished draft is somewhat different from what is found in
the official NAEP documentation (NCES, n.d.) or (AIR, n.d.) . Johnson
\& Rust (1992) found that, on average, the Satterthwaite approximation
underestimates the true DoF when $\nu_{k}$ are small and especially,
when we have $\nu_{k}=1$ for all $k$. Prominently, $\nu_{k}=1$
is the case in Jackknife variance estimation and balanced repeated
replicates (BRR) estimation of the variance. The adjustment suggested
by Johnson \& Rust (1992) was also studied by Qian (1998) and is given
by 
\[
\lambda_{J\&R}=\left(3.16-\frac{2.77}{\sqrt{K}}\right)
\]
where $K=62,$$\sqrt{62}=7.87$ (and in the Johnson \& Rust paper
$f$ is used rather than $\nu$). For NAEP $\left(M=\right)62=K$
and $\left(f=\right)1=\nu$ we have 
\[
\lambda_{J\&R}\approx2.808
\]

The simulation study reported by Johnson \& Rust (1992) produces a
table that summarizes the relationship between number of PSUs $K$,
degrees of freedom per term in the complex variance estimator $\nu$,
which equals 1 in the case of JRR, and the resulting true DoF $K\times\nu$
and the Satterthwaite approximate effective DoF in terms of median
and mean ratio to true DoF for this estimate. 

The table provided by Johnson \& Rust (1992) is reproduced in the
last section of this paper together with results that compare the
current NAEP adjustment baased on Johnson \& Rust (1992) and the newly
proposed adjustment based on a better approximation for small $K$
and $\nu$.

\section{A More General Estimate of the Degrees of Freedom}

Replacement of the variance with an estimate requires making certain
assumptions that we ignored - or at least not mentioned - above. A
different set of assumptions is needed in the case that $\nu_{k}$
are small or even $\nu_{k}=1$, and also for small $K$. Recall that
we obtained
\[
\nu_{?}=\frac{\left(\sigma_{*}^{2}\right)^{2}}{\sum_{k=1}^{K}\frac{\left(\sigma_{k}^{2}\right)^{2}}{\nu_{k}}}
\]

We still need to replace the unknown variance $\left(\sigma_{Q}^{2}\right)^{2}$
by an expression that uses the $S_{Q}^{2}$ but acknowledges that
$S_{Q}^{4}$ has a different expected value $E\left(S_{Q}^{4}\right)$for
small $\nu_{k}$ 
\[
\left(\sigma_{Q}^{2}\right)^{2}=\sigma_{Q}^{4}\approx R_{Q}\left(S_{Q}^{4}\right)
\]

We again require this for both cases $Q=k$ and $Q=*$. For moderate
and large $K$, we continue to replace 
\[
\left(\sigma_{*}^{2}\right)^{2}\approx\left(S_{*}^{2}\right)^{2}=\left(\sum_{k=1}^{K}S_{K}^{2}\right)^{2}.
\]
The squared expectation of a sum of $K$ independent terms equals
the square of the sum of expectations, that is, we have 
\[
\left[E\left(\sum_{i}X_{i}^{2}\right)\right]^{2}=\left[\sum_{i}E\left(X_{i}^{2}\right)\right]^{2}.
\]

\subsection{Important Special Case is $K=2$, Trivial Case is $K=1$}

Any approximation of degrees of freedom of a complex variance estimate
can assume that $K>1.$ This is easy to see by assuming $K=1.$ In
this case, we obtain
\[
\nu_{?}=\frac{\left(\sum_{k=1}^{1}\sigma_{k}^{2}\right)^{2}}{\sum_{k=1}^{1}\frac{\left(\sigma_{k}^{2}\right)^{2}}{\nu_{k}}}=\frac{\left(\sigma_{1}^{2}\right)^{2}}{\frac{\left(\sigma_{1}^{2}\right)^{2}}{\nu_{1}}}=\nu_{1}
\]

Therefore, while $K=1$ is technically possible, it is a trivial case
where the estimated effective d.f. are equal to the d.f. of the only
variance component in the expression. Obvioulsy, as Satterthwaite
(1941) was discussing, complex variance estimates, one can assume
that the trivial case was never intended to be considered. Hence,
we may assume only cases where $K>1$ holds.

The minimally useful case is the one where two variance components
are combined, namely
\[
\sigma^{2}=\sigma_{1}^{2}+\sigma_{2}^{2}.
\]
This case is of importance for the Welch (1947) improved t-test, as
well as the Rubin \& Schenker (1986) approach for d.f. of complex
estimates of variances involving imputations. Note that Lipsitz et
al. (2002) use the Welch-Satterthwaite equation to improve the Rubin\&
Schenker approach for the case of variance expressions that include
sampling based components with finite degrees of freedom. All of these,
Welch (1947), Rubin \& Schenker (1986), and Lipsitz (2002) can, in
essence, be considered straightforward applications of the Satterthwaite
approach for $K=2.$ 

In our special case, if both components are estimating the same variance
$\sigma^{2}$ with $\nu=1$ degrees of freedom we obtain
\[
E\left(\left[S_{1}^{2}+S_{2}^{2}\right]^{2}\right)=\sigma^{4}E\left(\left[Z_{1}^{2}+Z_{2}^{2}\right]^{2}\right)=\sigma^{4}K\left(2+K\right)=8\sigma^{4}
\]
so the expected value is not three times, (as for $K=1$ we have $1\left(2+1\right)$),
but eight times $\sigma^{4}.$

\subsection{A General Plug-In Estimator for $\sigma_{k}^{4}$}

When replacing 
\[
\left(\sigma_{k}^{2}\right)^{2}\approx R_{1k}\left(\left[S_{k}^{2}\right]^{2}\right)
\]

recall that 
\[
S_{k}^{2}=\frac{1}{\nu_{k}}\sum_{i=1}^{\nu_{k}+1}Z_{i}^{2}\sigma_{k}^{2}
\]
Then, we can us the second useful identity, which is 
\[
\nu_{k}^{2}E\left(\left[S_{k}^{2}\right]^{2}\right)=E\left(\left[\sum_{i}^{\nu_{k}}Z_{i}^{2}\sigma^{2}\right]^{2}\right)=\sigma_{k}^{4}\nu_{k}\left(\nu_{k}+2\right)
\]
Which yields
\[
\frac{E\left(\left[S_{k}^{2}\right]^{2}\right)}{\left(\nu_{k}+2\right)}=\frac{\left(\sigma_{k}^{2}\right)^{2}}{\nu_{k}}
\]

so that the appropriate replacement is 

\[
\sum_{k=1}^{K}\frac{\left(\sigma_{k}^{2}\right)^{2}}{\nu_{k}}\approx\sum_{k=1}^{K}\frac{\left(S_{k}^{2}\right)^{2}}{\nu_{k}+2}.
\]
 For $\nu_{k}=1$ for all $k$ this yields the following equation
\[
\nu_{?}\approx\hat{\nu}=\frac{\left(\sum_{k=1}^{K}\sigma_{k}^{2}\right)^{2}}{\sum_{k=1}^{K}\frac{\left(S_{k}^{2}\right)^{2}}{3}}=3\frac{\left(\sum_{k=1}^{K}\sigma_{k}^{2}\right)^{2}}{\sum_{k=1}^{K}\left(S_{k}^{2}\right)^{2}}
\]

and for general $\nu_{k}$ we note that as $\nu_{k}\rightarrow\infty$,
we obtain
\[
\nu_{?}\approx\hat{\nu}=\frac{\left(\sum_{k=1}^{K}\sigma_{k}^{2}\right)^{2}}{\sum_{k=1}^{K}\frac{\left(S_{k}^{2}\right)^{2}}{\nu_{k}+2}}\approx\frac{\left(\sum_{k=1}^{K}\sigma_{k}^{2}\right)^{2}}{\sum_{k=1}^{K}\frac{\left(S_{k}^{2}\right)^{2}}{\nu_{k}}}
\]

the next section derives the appropriate replacement for $\left(\sum_{k=1}^{K}\sigma_{k}^{2}\right)^{2}$.

\subsection{A General Plug-In Estimator for $\left(\sigma_{*}^{2}\right)^{2}$}

For the expression $\left(\sigma_{*}^{2}\right)^{2}=\left(\sum_{k=1}^{K}\sigma_{k}^{2}\right)^{2}$
we note, similar to the argument above, that if $K>1,$and $\nu_{k}=1,$and
with 
\[
E\left(S_{1}^{2}\right)=\sigma^{2},
\]
we have, implied by the second \emph{useful identity}, for $K>1$
and $\nu_{k}=1$, 

\[
E\left[\left(S_{*}^{2}\right)^{2}\right]=E\left[\left(\sum_{k=1}^{K}Z_{i}^{2}\sigma^{2}\right)^{2}\right]=\sigma^{4}K\left(K+2\right)
\]
while, for growing number of variance components, $K\rightarrow\infty$,
we may write 
\[
E\left[\left(S_{*}^{2}\right)^{2}\right]=\sigma^{4}E\left[\left(\sum_{k=1}^{K}Z_{k}^{2}\right)^{2}\right]\rightarrow\sigma^{4}K^{2}
\]
since $\frac{K^{2}}{K\left(K+2\right)}=\frac{K}{K+2}\rightarrow1$
for $K\rightarrow\infty.$ 

If $\nu_{k}=\nu>1$ is the same for all $k$ and with $\nu,K\rightarrow\infty$
we obtain
\[
E\left[\left(S_{*}^{2}\right)^{2}\right]=E\left[\left(\sum_{k=1}^{K}S_{k}^{2}\right)^{2}\right]=E\left[\left(\sum_{k=1}^{K}\frac{1}{\nu}\sum_{i=1}^{\nu+1}\left(x_{ik}-M_{k}\right)^{2}\right)^{2}\right]\rightarrow E\left[\left(K\sigma^{2}\right)^{2}\right]=\sigma^{4}K^{2}
\]
since $\frac{1}{\nu}\sum_{i=1}^{\nu+1}\left(x_{ik}-M_{k}\right)^{2}\rightarrow\sigma^{2}$
with $\nu\rightarrow\infty$. Finally, if $K=1$ and $\nu_{1}=\nu\rightarrow\infty$
we also obtain $E\left[\left(S_{*}^{2}\right)^{2}\right]=E\left[\left(\sigma^{2}\right)^{2}\right]=\sigma^{4}=\sigma^{4}K^{2}$
since $K=K^{2}=1$. 

Hence, there is need to select a $\lambda_{*}$so that 
\[
\lambda_{*}E\left[\left(S_{*}^{2}\right)^{2}\right]\approx\sigma^{4}K^{2}
\]
for small $K$ and $\nu_{k}$, as well either $K\rightarrow\infty$
growing or $\nu_{k}\rightarrow\infty$. This goal to produce a general
adjustment is served by using 
\[
\lambda_{*}=1+\frac{2}{\sum_{k=1}^{K}\nu_{k}}.
\]
For cases where $\nu_{k}=\nu$ for all $k$ we obtain $\lambda_{*}=1+\frac{2}{K\nu}$
and for $\nu=1$ we have $\lambda_{*}=1+\frac{2}{K}.$ For cases with
$\sum_{k}\nu_{k}\rightarrow\infty$ we have $\lambda_{*}=1+\frac{2}{\sum_{k}\nu_{k}}\rightarrow1$
and hence the use of $\lambda_{*}$ensures that.
\[
E\left[\frac{\left(\sum_{k=1}^{K}S_{k}^{2}\right)^{2}}{\left(1+\frac{2}{\sum_{k}\nu_{k}}\right)}\right]\approx\left(\sum_{k=1}^{K}\sigma_{k}^{2}\right)^{2}=\sigma^{4}K^{2}.
\]

The next section shows how utilizing the two terms to approximate
$\left(\sigma_{*}^{2}\right)^{2}=\left(\sum_{k=1}^{K}\sigma_{k}^{2}\right)^{2}$
and the $\left(\sigma_{k}^{2}\right)^{2}$ leads to an estimator of
the effective degrees of freedom that yields the expected result even
for small degrees of freedom $\nu_{k}<<\infty$ including the important
special case of $\nu_{k}=\nu=1$ for all $k.$

\subsection{An Improved Effective Degrees of Freedom Estimator}

With both adjustments derived above to match the expected values,
we obtain an expression that is close to the Johnson \& Rust (1992)
adjustment, but has a theoretical rather than empirical rationale
grounded in the espectation f powers of normally distributed variables.
The proposed adjustment is more general than the empirical adjustment
by Johnson \& Rust, which was derived based on a simulation result
designed for NAEP. The proposed adjustment also works in cases other
than the $\nu_{k}=1$ case, and provides a vanishing adjustment as
the $K,\nu_{k}$ increase. 

The proposed estimator of effective degrees of freedom becomes

\[
\nu_{?}\approx\hat{\nu}=\frac{\left(\sum_{k=1}^{K}S_{k}^{2}\right)^{2}}{\left(1+\frac{2}{\sum_{k}\nu_{k}}\right)\left(\sum_{k=1}^{K}\frac{\left(S_{k}^{2}\right)^{2}}{\nu_{k}+2}\right)}
\]
or
\begin{equation}
\nu_{?}\approx\hat{\nu}=\frac{\left(\nu+2\right)\left(\sum_{k=1}^{K}S_{k}^{2}\right)^{2}}{\left(1+\frac{2}{K\nu}\right)\left(\sum_{k=1}^{K}\left(S_{k}^{2}\right)^{2}\right)}=\frac{\left(\sum_{k=1}^{K}S_{k}^{2}\right)^{2}}{\left(\sum_{k=1}^{K}\left(S_{k}^{2}\right)^{2}\right)}A\left(\nu,K\right)\label{eq:adjust1}
\end{equation}

for cases where $\nu_{k}=\nu$ for all $k=1,...,K.$ That ism in the
case of $\nu_{k}=1$ for all $k,$the proposed adjustment is obtained
by multiplying the original estimator by 
\[
A\left(\nu,K\right)=\frac{\left(\nu+2\right)}{\left(1+\frac{2}{K\nu}\right)}.
\]

\begin{table}[h]
\selectlanguage{english}%
\begin{tabular}{|c|c|c|c|c|c|c|c|c|}
\hline 
\foreignlanguage{american}{{\footnotesize ($K$)}} & \foreignlanguage{american}{{\footnotesize ($\nu$)}} & \foreignlanguage{american}{{\footnotesize$K\nu$}} & {\footnotesize Mean} & {\footnotesize Median} & {\footnotesize Upper} & {\footnotesize Lower} & {\footnotesize Proposed} & {\footnotesize NAEP}\tabularnewline
{\footnotesize M} & {\footnotesize f} & {\footnotesize dftrue} & {\footnotesize Ratio} & {\footnotesize Ratio} & {\footnotesize Quartile} & {\footnotesize Quartile} & {\footnotesize Adjustment} & {\footnotesize Adjustment}\tabularnewline
\hline 
{\footnotesize 5} & {\footnotesize 1} & {\footnotesize 5} & {\footnotesize 0.51} & {\footnotesize 0.5} & {\footnotesize 0.61} & {\footnotesize 0.4} & {\footnotesize 1.093} & {\footnotesize 0.980}\tabularnewline
{\footnotesize 10} & {\footnotesize 1} & {\footnotesize 10} & {\footnotesize 0.43} & {\footnotesize 0.43} & {\footnotesize 0.52} & {\footnotesize 0.35} & {\footnotesize 1.075} & {\footnotesize 0.982}\tabularnewline
{\footnotesize 20} & {\footnotesize 1} & {\footnotesize 20} & {\footnotesize 0.39} & {\footnotesize 0.39} & {\footnotesize 0.45} & {\footnotesize 0.32} & {\footnotesize 1.064} & {\footnotesize 0.991}\tabularnewline
{\footnotesize 30} & {\footnotesize 1} & {\footnotesize 30} & {\footnotesize 0.38} & {\footnotesize 0.38} & {\footnotesize 0.43} & {\footnotesize 0.32} & {\footnotesize 1.069} & {\footnotesize 1.009}\tabularnewline
{\footnotesize 40} & {\footnotesize 1} & {\footnotesize 40} & {\footnotesize 0.37} & {\footnotesize 0.37} & {\footnotesize 0.42} & {\footnotesize 0.32} & {\footnotesize 1.057} & {\footnotesize 1.007}\tabularnewline
{\footnotesize 50} & {\footnotesize 1} & {\footnotesize 50} & {\footnotesize 0.36} & {\footnotesize 0.36} & {\footnotesize 0.4} & {\footnotesize 0.31} & {\footnotesize 1.038} & {\footnotesize 0.997}\tabularnewline
{\footnotesize 100} & {\footnotesize 1} & {\footnotesize 100} & {\footnotesize 0.35} & {\footnotesize 0.35} & {\footnotesize 0.38} & {\footnotesize 0.31} & {\footnotesize 1.029} & {\footnotesize 1.009}\tabularnewline
{\footnotesize 5} & {\footnotesize 2} & {\footnotesize 10} & {\footnotesize 0.64} & {\footnotesize 0.65} & {\footnotesize 0.74} & {\footnotesize 0.54} & {\footnotesize 1.067} & {\footnotesize 1.230}\tabularnewline
{\footnotesize 10} & {\footnotesize 2} & {\footnotesize 20} & {\footnotesize 0.57} & {\footnotesize 0.58} & {\footnotesize 0.66} & {\footnotesize 0.5} & {\footnotesize 1.036} & {\footnotesize 1.302}\tabularnewline
{\footnotesize 20} & {\footnotesize 2} & {\footnotesize 40} & {\footnotesize 0.55} & {\footnotesize 0.55} & {\footnotesize 0.6} & {\footnotesize 0.48} & {\footnotesize 1.048} & {\footnotesize 1.397}\tabularnewline
{\footnotesize 30} & {\footnotesize 2} & {\footnotesize 60} & {\footnotesize 0.53} & {\footnotesize 0.53} & {\footnotesize 0.58} & {\footnotesize 0.48} & {\footnotesize 1.026} & {\footnotesize 1.407}\tabularnewline
{\footnotesize 40} & {\footnotesize 2} & {\footnotesize 80} & {\footnotesize 0.52} & {\footnotesize 0.53} & {\footnotesize 0.57} & {\footnotesize 0.48} & {\footnotesize 1.015} & {\footnotesize 1.415}\tabularnewline
{\footnotesize 50} & {\footnotesize 2} & {\footnotesize 100} & {\footnotesize 0.52} & {\footnotesize 0.52} & {\footnotesize 0.57} & {\footnotesize 0.48} & {\footnotesize 1.020} & {\footnotesize 1.439}\tabularnewline
{\footnotesize 100} & {\footnotesize 2} & {\footnotesize 200} & {\footnotesize 0.51} & {\footnotesize 0.51} & {\footnotesize 0.54} & {\footnotesize 0.48} & {\footnotesize 1.010} & {\footnotesize 1.470}\tabularnewline
{\footnotesize 5} & {\footnotesize 3} & {\footnotesize 15} & {\footnotesize 0.72} & {\footnotesize 0.73} & {\footnotesize 0.81} & {\footnotesize 0.63} & {\footnotesize 1.059} & {\footnotesize 1.383}\tabularnewline
{\footnotesize 10} & {\footnotesize 3} & {\footnotesize 30} & {\footnotesize 0.66} & {\footnotesize 0.66} & {\footnotesize 0.74} & {\footnotesize 0.59} & {\footnotesize 1.031} & {\footnotesize 1.507}\tabularnewline
{\footnotesize 20} & {\footnotesize 3} & {\footnotesize 60} & {\footnotesize 0.63} & {\footnotesize 0.63} & {\footnotesize 0.69} & {\footnotesize 0.57} & {\footnotesize 1.016} & {\footnotesize 1.601}\tabularnewline
{\footnotesize 30} & {\footnotesize 3} & {\footnotesize 90} & {\footnotesize 0.62} & {\footnotesize 0.63} & {\footnotesize 0.67} & {\footnotesize 0.58} & {\footnotesize 1.011} & {\footnotesize 1.646}\tabularnewline
{\footnotesize 40} & {\footnotesize 3} & {\footnotesize 120} & {\footnotesize 0.62} & {\footnotesize 0.62} & {\footnotesize 0.66} & {\footnotesize 0.58} & {\footnotesize 1.016} & {\footnotesize 1.688}\tabularnewline
{\footnotesize 50} & {\footnotesize 3} & {\footnotesize 150} & {\footnotesize 0.61} & {\footnotesize 0.61} & {\footnotesize 0.65} & {\footnotesize 0.57} & {\footnotesize 1.003} & {\footnotesize 1.689}\tabularnewline
{\footnotesize 100} & {\footnotesize 3} & {\footnotesize 300} & {\footnotesize 0.61} & {\footnotesize 0.61} & {\footnotesize 0.64} & {\footnotesize 0.58} & {\footnotesize 1.010} & {\footnotesize 1.759}\tabularnewline
{\footnotesize 5} & {\footnotesize 4} & {\footnotesize 20} & {\footnotesize 0.76} & {\footnotesize 0.77} & {\footnotesize 0.85} & {\footnotesize 0.68} & {\footnotesize 1.036} & {\footnotesize 1.460}\tabularnewline
{\footnotesize 10} & {\footnotesize 4} & {\footnotesize 40} & {\footnotesize 0.71} & {\footnotesize 0.72} & {\footnotesize 0.78} & {\footnotesize 0.65} & {\footnotesize 1.014} & {\footnotesize 1.622}\tabularnewline
{\footnotesize 20} & {\footnotesize 4} & {\footnotesize 80} & {\footnotesize 0.7} & {\footnotesize 0.7} & {\footnotesize 0.75} & {\footnotesize 0.65} & {\footnotesize 1.024} & {\footnotesize 1.778}\tabularnewline
{\footnotesize 30} & {\footnotesize 4} & {\footnotesize 120} & {\footnotesize 0.69} & {\footnotesize 0.69} & {\footnotesize 0.73} & {\footnotesize 0.65} & {\footnotesize 1.018} & {\footnotesize 1.831}\tabularnewline
{\footnotesize 40} & {\footnotesize 4} & {\footnotesize 160} & {\footnotesize 0.68} & {\footnotesize 0.68} & {\footnotesize 0.72} & {\footnotesize 0.65} & {\footnotesize 1.007} & {\footnotesize 1.851}\tabularnewline
{\footnotesize 50} & {\footnotesize 4} & {\footnotesize 200} & {\footnotesize 0.68} & {\footnotesize 0.68} & {\footnotesize 0.71} & {\footnotesize 0.65} & {\footnotesize 1.010} & {\footnotesize 1.882}\tabularnewline
{\footnotesize 100} & {\footnotesize 4} & {\footnotesize 400} & {\footnotesize 0.67} & {\footnotesize 0.67} & {\footnotesize 0.7} & {\footnotesize 0.65} & {\footnotesize 1.000} & {\footnotesize 1.932}\tabularnewline
{\footnotesize 5} & {\footnotesize 5} & {\footnotesize 25} & {\footnotesize 0.79} & {\footnotesize 0.8} & {\footnotesize 0.87} & {\footnotesize 0.72} & {\footnotesize 1.024} & {\footnotesize 1.518}\tabularnewline
{\footnotesize 10} & {\footnotesize 5} & {\footnotesize 50} & {\footnotesize 0.76} & {\footnotesize 0.76} & {\footnotesize 0.82} & {\footnotesize 0.7} & {\footnotesize 1.023} & {\footnotesize 1.736}\tabularnewline
{\footnotesize 20} & {\footnotesize 5} & {\footnotesize 100} & {\footnotesize 0.73} & {\footnotesize 0.74} & {\footnotesize 0.78} & {\footnotesize 0.69} & {\footnotesize 1.002} & {\footnotesize 1.855}\tabularnewline
{\footnotesize 30} & {\footnotesize 5} & {\footnotesize 150} & {\footnotesize 0.73} & {\footnotesize 0.73} & {\footnotesize 0.77} & {\footnotesize 0.69} & {\footnotesize 1.009} & {\footnotesize 1.938}\tabularnewline
{\footnotesize 40} & {\footnotesize 5} & {\footnotesize 200} & {\footnotesize 0.73} & {\footnotesize 0.73} & {\footnotesize 0.76} & {\footnotesize 0.7} & {\footnotesize 1.012} & {\footnotesize 1.987}\tabularnewline
{\footnotesize 50} & {\footnotesize 5} & {\footnotesize 250} & {\footnotesize 0.72} & {\footnotesize 0.72} & {\footnotesize 0.75} & {\footnotesize 0.69} & {\footnotesize 1.000} & {\footnotesize 1.993}\tabularnewline
{\footnotesize 100} & {\footnotesize 5} & {\footnotesize 500} & {\footnotesize 0.72} & {\footnotesize 0.72} & {\footnotesize 0.74} & {\footnotesize 0.7} & {\footnotesize 1.004} & {\footnotesize 2.076}\tabularnewline
{\footnotesize 5} & {\footnotesize 10} & {\footnotesize 50} & {\footnotesize 0.87} & {\footnotesize 0.88} & {\footnotesize 0.93} & {\footnotesize 0.83} & {\footnotesize 1.004} & {\footnotesize 1.671}\tabularnewline
{\footnotesize 10} & {\footnotesize 10} & {\footnotesize 100} & {\footnotesize 0.85} & {\footnotesize 0.86} & {\footnotesize 0.89} & {\footnotesize 0.82} & {\footnotesize 1.000} & {\footnotesize 1.941}\tabularnewline
{\footnotesize 20} & {\footnotesize 10} & {\footnotesize 200} & {\footnotesize 0.85} & {\footnotesize 0.85} & {\footnotesize 0.88} & {\footnotesize 0.82} & {\footnotesize 1.010} & {\footnotesize 2.160}\tabularnewline
{\footnotesize 30} & {\footnotesize 10} & {\footnotesize 300} & {\footnotesize 0.84} & {\footnotesize 0.84} & {\footnotesize 0.87} & {\footnotesize 0.82} & {\footnotesize 1.001} & {\footnotesize 2.230}\tabularnewline
{\footnotesize 40} & {\footnotesize 10} & {\footnotesize 400} & {\footnotesize 0.84} & {\footnotesize 0.84} & {\footnotesize 0.86} & {\footnotesize 0.82} & {\footnotesize 1.003} & {\footnotesize 2.287}\tabularnewline
{\footnotesize 50} & {\footnotesize 10} & {\footnotesize 500} & {\footnotesize 0.84} & {\footnotesize 0.84} & {\footnotesize 0.86} & {\footnotesize 0.82} & {\footnotesize 1.004} & {\footnotesize 2.325}\tabularnewline
{\footnotesize 100} & {\footnotesize 10} & {\footnotesize 1000} & {\footnotesize 0.84} & {\footnotesize 0.84} & {\footnotesize 0.85} & {\footnotesize 0.82} & {\footnotesize 1.006} & {\footnotesize 2.422}\tabularnewline
{\footnotesize 5} & {\footnotesize 25} & {\footnotesize 125} & {\footnotesize 0.94} & {\footnotesize 0.95} & {\footnotesize 0.97} & {\footnotesize 0.92} & {\footnotesize 0.999} & {\footnotesize 1.806}\tabularnewline
{\footnotesize 10} & {\footnotesize 25} & {\footnotesize 250} & {\footnotesize 0.94} & {\footnotesize 0.94} & {\footnotesize 0.96} & {\footnotesize 0.92} & {\footnotesize 1.007} & {\footnotesize 2.147}\tabularnewline
{\footnotesize 20} & {\footnotesize 25} & {\footnotesize 500} & {\footnotesize 0.93} & {\footnotesize 0.93} & {\footnotesize 0.95} & {\footnotesize 0.92} & {\footnotesize 1.000} & {\footnotesize 2.363}\tabularnewline
{\footnotesize 30} & {\footnotesize 25} & {\footnotesize 750} & {\footnotesize 0.93} & {\footnotesize 0.93} & {\footnotesize 0.94} & {\footnotesize 0.92} & {\footnotesize 1.002} & {\footnotesize 2.468}\tabularnewline
{\footnotesize 40} & {\footnotesize 25} & {\footnotesize 1000} & {\footnotesize 0.93} & {\footnotesize 0.93} & {\footnotesize 0.94} & {\footnotesize 0.92} & {\footnotesize 1.002} & {\footnotesize 2.531}\tabularnewline
{\footnotesize 50} & {\footnotesize 25} & {\footnotesize 1250} & {\footnotesize 0.93} & {\footnotesize 0.93} & {\footnotesize 0.94} & {\footnotesize 0.92} & {\footnotesize 1.003} & {\footnotesize 2.574}\tabularnewline
{\footnotesize 100} & {\footnotesize 25} & {\footnotesize 2500} & {\footnotesize 0.93} & {\footnotesize 0.93} & {\footnotesize 0.93} & {\footnotesize 0.92} & {\footnotesize 1.004} & {\footnotesize 2.681}\tabularnewline
\hline 
\end{tabular}\caption{The Johnson \& Rust (1992) simulation results for Satterthwaite's
(1946) effective degrees of freedom with proposed and current NAEP
adjustment.}
\foreignlanguage{american}{\label{j_and_r_table1}}\selectlanguage{american}%
\end{table}

The results given in the second to last column show how the proposed
adjustment removes the downward bias for all $K=M,\nu=f$ cases examined
in the Johnson \& Rust (1992) simulation, and it provides an estimate
that is slightly larger than 1 for small $K$ and $\nu$ while it
approaches 1.0 for increasing values. 

In contrast, the approach used in NAEP (AIR, n.d., NCES, n.d.) based
on the Johnson \& Rust (1992) simulation was optimized for the NAEP
case, and only works appropriately for cases where $\nu=1$ and overestimates
the DoF with growing $K$ and $\nu$. 

\subsection{Further Exploration and Improvement of the Estimator}

Note that for $K=1$ the proposed adjusted formula results in

\[
\nu_{?}\approx\hat{\nu}=\frac{\left(S_{1}^{2}\right)^{2}}{\frac{1}{\nu+2}\left(1+\frac{2}{K\nu}\right)\left(S_{1}^{2}\right)^{2}}=\frac{\nu+2}{1+\frac{2}{K\nu}}=\frac{\nu+2}{1+\frac{2}{\nu}}
\]
and further 
\[
\frac{\nu+2}{1+\frac{2}{\nu}}=\frac{\nu+2}{\frac{1}{\nu}\left(\nu+2\right)}=\nu
\]

One can argue that $K=1$ is a trivial case, where the Satterthwaite
estimate is not necessary because the result for $K=1$ is constant
$\nu$: Here we do not need an estimate of the effective DoF, we already
know the result is $\nu_{?}=\nu_{1}=\nu$. Therefore, we can limit
our considerations to cases where $K\ge2$. Define the average of
the variance components as
\[
\overline{S^{2}}=\sum_{k=1}^{K}S_{k}^{2}
\]
 and 
\[
R_{K}^{2}=\frac{S_{k}^{2}}{\overline{S^{2}}}.
\]
 Then we have 
\[
\nu_{?}\approx\hat{\nu}=\frac{\left(\nu+2\right)\left(\overline{S^{2}}\sum_{k=1}^{K}\frac{S_{k}^{2}}{\overline{S^{2}}}\right)^{2}}{\left(1+\frac{2}{K\nu}\right)\left(\overline{S^{2}}^{2}\sum_{k=1}^{K}\left(\frac{S_{k}^{2}}{\overline{S^{2}}}\right)^{2}\right)}=\frac{\left(\nu+2\right)\left(\sum_{k=1}^{K}R_{K}^{2}\right)^{2}}{\left(1+\frac{2}{K\nu}\right)\left(\sum_{k=1}^{K}\left(R_{K}^{2}\right)^{2}\right)}
\]

It is easy to show that 
\[
\left(\sum_{k=1}^{K}R_{K}^{2}\right)^{2}=\left(\sum_{k=1}^{K}\frac{S_{k}^{2}}{\overline{S^{2}}}\right)^{2}=\left(\frac{1}{\overline{S^{2}}}\left[\sum_{k=1}^{K}S_{k}^{2}\right]\right)^{2}=\left(\frac{1}{\overline{S^{2}}}\left[\overline{S^{2}}\right]\right)^{2}=1.
\]
However, for the sum of the squared terms we obtain
\[
\left(\sum_{k=1}^{K}\left(R_{K}^{2}\right)^{2}\right)=\left(\sum_{k=1}^{K}\left(\frac{S_{k}^{2}}{\overline{S^{2}}}\right)^{2}\right)=\left[\frac{1}{\overline{S^{2}}}\right]^{2}\left(\sum_{k=1}^{K}S_{k}^{4}\right)
\]

for $K=2$ we have 
\[
\overline{S^{2}}=\left(S_{1}^{2}+S_{2}^{2}\right)
\]
and 
\[
\left[\overline{S^{2}}\right]^{2}=\left(S_{1}^{2}+S_{2}^{2}\right)^{2}
\]
so that we obtain
\[
\left(\sum_{k=1}^{2}\left(R_{K}^{2}\right)^{2}\right)=\frac{\left(S_{1}^{4}+S_{2}^{4}\right)}{\left(S_{1}^{2}+S_{2}^{2}\right)^{2}}
\]

if $S_{1}^{2}=S_{2}^{2}$ we obtain
\[
\frac{\left(S_{1}^{4}+S_{2}^{4}\right)}{\left(S_{1}^{2}+S_{2}^{2}\right)^{2}}=\frac{\left(S_{1}^{4}+S_{1}^{4}\right)}{\left(S_{1}^{2}+S_{1}^{2}\right)^{2}}=\frac{2S_{1}^{4}}{4\left(S_{1}^{2}\right)^{2}}=\frac{1}{2}
\]
as expected. If either one of them is zero, say $S_{1}^{2}=0,S_{2}^{2}>0$,
then we have
\[
\frac{\left(S_{1}^{4}+S_{2}^{4}\right)}{\left(S_{1}^{2}+S_{2}^{2}\right)^{2}}=\frac{\left(S_{2}^{4}\right)}{\left(S_{2}^{2}\right)^{2}}=1
\]

For $\nu=1$ and $K=2$ we obtain
\[
\hat{\nu}=\frac{\left(\nu+2\right)1}{\left(1+\frac{2}{K\nu}\right)\left(\sum_{k=1}^{K}\left(R_{K}^{2}\right)^{2}\right)}=\frac{3}{2\left(\sum_{k=1}^{K}\left(R_{K}^{2}\right)^{2}\right)}=\frac{3/2}{\left(\sum_{k=1}^{K}\left(R_{K}^{2}\right)^{2}\right)}
\]

if $S_{1}^{2}=0,S_{2}^{2}>0$ we have $\left(\sum_{k=1}^{K}\left(R_{K}^{2}\right)^{2}\right)=1$
so the result is
\[
\hat{\nu}=\frac{3}{2}=1.5>1
\]

and if $S_{1}^{2}=S_{2}^{2}$ we have $\left(\sum_{k=1}^{K}\left(R_{K}^{2}\right)^{2}\right)=\frac{1}{2}$
so that
\[
\hat{\nu}=\frac{3}{2}/\frac{1}{2}=3>2
\]
which means that the proposed adjustment given in \ref{eq:adjust1}
\emph{over-adjusts} in the case of $K=2$. Therefore, a modified adjustment
is defined with
\[
\lambda_{*}=1+\frac{2}{\left(K-1\right)\nu}.
\]
This adjustment results in 
\[
\hat{\nu}=\frac{\left(\nu+2\right)K^{2}}{\left(1+\frac{2}{(K-1)\nu}\right)\left(\sum_{k=1}^{K}\left(R_{K}^{2}\right)^{2}\right)}=\frac{12}{\left(1+\frac{2}{(2-1)1}\right)\left(\sum_{k=1}^{K}\left(R_{K}^{2}\right)^{2}\right)}=\frac{4}{\left(\sum_{k=1}^{K}\left(R_{K}^{2}\right)^{2}\right)}
\]

With $S_{1}^{2}=0,S_{2}^{2}>0$ we have $\left(\sum_{k=1}^{K}\left(R_{K}^{2}\right)^{2}\right)=4$
so that, as expected,
\[
\hat{\nu}=\frac{4}{4}=1,
\]

and with $S_{1}^{2}=S_{2}^{2}$ we have $\left(\sum_{k=1}^{K}\left(R_{K}^{2}\right)^{2}\right)=2$
so that, as expected, 
\[
\hat{\nu}=\frac{4}{2}=2=K.
\]

\subsection{A Closer Approximation of the Effective DoF}

The result derived in the above section leads to a closer approximation
of the effective DoF. Given the above result it is proposed to use
\[
\lambda_{*}=1+\frac{2}{\left(1-\frac{1}{K}\right)\sum_{k}\nu_{k}}
\]
or for cases where all $\nu_{k}=\nu,$we can write
\[
\lambda_{*}=1+\frac{2}{\left(K-1\right)\nu}
\]
For $K=2,\nu=1$ we obtain the largest adjustment 
\[
\lambda_{*}=3
\]
and, as before, as either $K\rightarrow\infty$ or $\nu\rightarrow\infty$,
we continue to obtain the same limit of 
\[
\lambda_{*}=1
\]

so we approach asymptotically the original Satterthwaite (1946) expression
as the degrees of freedom per variance component $\nu_{k}$and the
number of components $K$ grow. This closer bound is therefore given
by 
\begin{equation}
\hat{\nu}=\frac{\left(\sum_{k=1}^{K}S_{k}^{2}\right)^{2}}{\left(1+\frac{2}{\left(1-\frac{1}{K}\right)\sum_{k}\nu_{k}}\right)\left(\sum_{k=1}^{K}\frac{\left(S_{k}^{2}\right)^{2}}{\nu_{k}+2}\right)}\label{eq:adjust_fin_unequal}
\end{equation}
 or if $\forall k:\nu_{k}=\nu$
\begin{equation}
\hat{\nu}=\frac{\left(\nu+2\right)\left(\sum_{k=1}^{K}S_{k}^{2}\right)^{2}}{\left(1+\frac{2}{\left(K-1\right)\nu}\right)\left(\sum_{k=1}^{K}\left(S_{k}^{2}\right)^{2}\right)}\label{eq:adjust_fin_equal}
\end{equation}

The resulting expected effective degrees of freedom estimated by either
\ref{eq:adjust_fin_unequal} or \ref{eq:adjust_fin_equal} are closely
tracking the true degrees of freedom as shown in Table \ref{j_and_r_table1-1}.
It is therefore suggested to use these adjusted estimates instead
of the original Satterthwaite (1941, 1946) formula, or the empirically
derived correction by Johnson \& Rust. The result presented here has
wide application and is not limited to cases where $\nu_{k}=1$, and
hence applies both to replication based variance estimators as well
as other complex (Satterthwaite, 1941) estimates of variance.

\begin{table}[h]
\selectlanguage{english}%
\begin{tabular}{|c|c|c|c|c|c|c|c|c|}
\hline 
\foreignlanguage{american}{{\footnotesize ($K$)}} & \foreignlanguage{american}{{\footnotesize ($\nu$)}} & \foreignlanguage{american}{{\footnotesize$K\nu$}} & {\footnotesize Mean} & {\footnotesize Median} & {\footnotesize Upper} & {\footnotesize Lower} & {\footnotesize Improved} & {\footnotesize NAEP}\tabularnewline
{\footnotesize M} & {\footnotesize f} & {\footnotesize dftrue} & {\footnotesize Ratio} & {\footnotesize Ratio} & {\footnotesize Quartile} & {\footnotesize Quartile} & {\footnotesize Adjustment} & {\footnotesize Adjustment}\tabularnewline
\hline 
{\footnotesize 5} & {\footnotesize 1} & {\footnotesize 5} & {\footnotesize 0.51} & {\footnotesize 0.5} & {\footnotesize 0.61} & {\footnotesize 0.4} & {\footnotesize 1.020} & {\footnotesize 0.980}\tabularnewline
{\footnotesize 10} & {\footnotesize 1} & {\footnotesize 10} & {\footnotesize 0.43} & {\footnotesize 0.43} & {\footnotesize 0.52} & {\footnotesize 0.35} & {\footnotesize 1.055} & {\footnotesize 0.982}\tabularnewline
{\footnotesize 20} & {\footnotesize 1} & {\footnotesize 20} & {\footnotesize 0.39} & {\footnotesize 0.39} & {\footnotesize 0.45} & {\footnotesize 0.32} & {\footnotesize 1.059} & {\footnotesize 0.991}\tabularnewline
{\footnotesize 30} & {\footnotesize 1} & {\footnotesize 30} & {\footnotesize 0.38} & {\footnotesize 0.38} & {\footnotesize 0.43} & {\footnotesize 0.32} & {\footnotesize 1.066} & {\footnotesize 1.009}\tabularnewline
{\footnotesize 40} & {\footnotesize 1} & {\footnotesize 40} & {\footnotesize 0.37} & {\footnotesize 0.37} & {\footnotesize 0.42} & {\footnotesize 0.32} & {\footnotesize 1.056} & {\footnotesize 1.007}\tabularnewline
{\footnotesize 50} & {\footnotesize 1} & {\footnotesize 50} & {\footnotesize 0.36} & {\footnotesize 0.36} & {\footnotesize 0.4} & {\footnotesize 0.31} & {\footnotesize 1.038} & {\footnotesize 0.997}\tabularnewline
{\footnotesize 100} & {\footnotesize 1} & {\footnotesize 100} & {\footnotesize 0.35} & {\footnotesize 0.35} & {\footnotesize 0.38} & {\footnotesize 0.31} & {\footnotesize 1.029} & {\footnotesize 1.009}\tabularnewline
{\footnotesize 5} & {\footnotesize 2} & {\footnotesize 10} & {\footnotesize 0.64} & {\footnotesize 0.65} & {\footnotesize 0.74} & {\footnotesize 0.54} & {\footnotesize 1.024} & {\footnotesize 1.230}\tabularnewline
{\footnotesize 10} & {\footnotesize 2} & {\footnotesize 20} & {\footnotesize 0.57} & {\footnotesize 0.58} & {\footnotesize 0.66} & {\footnotesize 0.5} & {\footnotesize 1.026} & {\footnotesize 1.302}\tabularnewline
{\footnotesize 20} & {\footnotesize 2} & {\footnotesize 40} & {\footnotesize 0.55} & {\footnotesize 0.55} & {\footnotesize 0.6} & {\footnotesize 0.48} & {\footnotesize 1.045} & {\footnotesize 1.397}\tabularnewline
{\footnotesize 30} & {\footnotesize 2} & {\footnotesize 60} & {\footnotesize 0.53} & {\footnotesize 0.53} & {\footnotesize 0.58} & {\footnotesize 0.48} & {\footnotesize 1.025} & {\footnotesize 1.407}\tabularnewline
{\footnotesize 40} & {\footnotesize 2} & {\footnotesize 80} & {\footnotesize 0.52} & {\footnotesize 0.53} & {\footnotesize 0.57} & {\footnotesize 0.48} & {\footnotesize 1.014} & {\footnotesize 1.415}\tabularnewline
{\footnotesize 50} & {\footnotesize 2} & {\footnotesize 100} & {\footnotesize 0.52} & {\footnotesize 0.52} & {\footnotesize 0.57} & {\footnotesize 0.48} & {\footnotesize 1.019} & {\footnotesize 1.439}\tabularnewline
{\footnotesize 100} & {\footnotesize 2} & {\footnotesize 200} & {\footnotesize 0.51} & {\footnotesize 0.51} & {\footnotesize 0.54} & {\footnotesize 0.48} & {\footnotesize 1.010} & {\footnotesize 1.470}\tabularnewline
{\footnotesize 5} & {\footnotesize 3} & {\footnotesize 15} & {\footnotesize 0.72} & {\footnotesize 0.73} & {\footnotesize 0.81} & {\footnotesize 0.63} & {\footnotesize 1.029} & {\footnotesize 1.383}\tabularnewline
{\footnotesize 10} & {\footnotesize 3} & {\footnotesize 30} & {\footnotesize 0.66} & {\footnotesize 0.66} & {\footnotesize 0.74} & {\footnotesize 0.59} & {\footnotesize 1.024} & {\footnotesize 1.507}\tabularnewline
{\footnotesize 20} & {\footnotesize 3} & {\footnotesize 60} & {\footnotesize 0.63} & {\footnotesize 0.63} & {\footnotesize 0.69} & {\footnotesize 0.57} & {\footnotesize 1.014} & {\footnotesize 1.601}\tabularnewline
{\footnotesize 30} & {\footnotesize 3} & {\footnotesize 90} & {\footnotesize 0.62} & {\footnotesize 0.63} & {\footnotesize 0.67} & {\footnotesize 0.58} & {\footnotesize 1.010} & {\footnotesize 1.646}\tabularnewline
{\footnotesize 40} & {\footnotesize 3} & {\footnotesize 120} & {\footnotesize 0.62} & {\footnotesize 0.62} & {\footnotesize 0.66} & {\footnotesize 0.58} & {\footnotesize 1.016} & {\footnotesize 1.688}\tabularnewline
{\footnotesize 50} & {\footnotesize 3} & {\footnotesize 150} & {\footnotesize 0.61} & {\footnotesize 0.61} & {\footnotesize 0.65} & {\footnotesize 0.57} & {\footnotesize 1.003} & {\footnotesize 1.689}\tabularnewline
{\footnotesize 100} & {\footnotesize 3} & {\footnotesize 300} & {\footnotesize 0.61} & {\footnotesize 0.61} & {\footnotesize 0.64} & {\footnotesize 0.58} & {\footnotesize 1.010} & {\footnotesize 1.759}\tabularnewline
{\footnotesize 5} & {\footnotesize 4} & {\footnotesize 20} & {\footnotesize 0.76} & {\footnotesize 0.77} & {\footnotesize 0.85} & {\footnotesize 0.68} & {\footnotesize 1.013} & {\footnotesize 1.460}\tabularnewline
{\footnotesize 10} & {\footnotesize 4} & {\footnotesize 40} & {\footnotesize 0.71} & {\footnotesize 0.72} & {\footnotesize 0.78} & {\footnotesize 0.65} & {\footnotesize 1.009} & {\footnotesize 1.622}\tabularnewline
{\footnotesize 20} & {\footnotesize 4} & {\footnotesize 80} & {\footnotesize 0.7} & {\footnotesize 0.7} & {\footnotesize 0.75} & {\footnotesize 0.65} & {\footnotesize 1.023} & {\footnotesize 1.778}\tabularnewline
{\footnotesize 30} & {\footnotesize 4} & {\footnotesize 120} & {\footnotesize 0.69} & {\footnotesize 0.69} & {\footnotesize 0.73} & {\footnotesize 0.65} & {\footnotesize 1.017} & {\footnotesize 1.831}\tabularnewline
{\footnotesize 40} & {\footnotesize 4} & {\footnotesize 160} & {\footnotesize 0.68} & {\footnotesize 0.68} & {\footnotesize 0.72} & {\footnotesize 0.65} & {\footnotesize 1.007} & {\footnotesize 1.851}\tabularnewline
{\footnotesize 50} & {\footnotesize 4} & {\footnotesize 200} & {\footnotesize 0.68} & {\footnotesize 0.68} & {\footnotesize 0.71} & {\footnotesize 0.65} & {\footnotesize 1.010} & {\footnotesize 1.882}\tabularnewline
{\footnotesize 100} & {\footnotesize 4} & {\footnotesize 400} & {\footnotesize 0.67} & {\footnotesize 0.67} & {\footnotesize 0.7} & {\footnotesize 0.65} & {\footnotesize 1.000} & {\footnotesize 1.932}\tabularnewline
{\footnotesize 5} & {\footnotesize 5} & {\footnotesize 25} & {\footnotesize 0.79} & {\footnotesize 0.8} & {\footnotesize 0.87} & {\footnotesize 0.72} & {\footnotesize 1.005} & {\footnotesize 1.518}\tabularnewline
{\footnotesize 10} & {\footnotesize 5} & {\footnotesize 50} & {\footnotesize 0.76} & {\footnotesize 0.76} & {\footnotesize 0.82} & {\footnotesize 0.7} & {\footnotesize 1.019} & {\footnotesize 1.736}\tabularnewline
{\footnotesize 20} & {\footnotesize 5} & {\footnotesize 100} & {\footnotesize 0.73} & {\footnotesize 0.74} & {\footnotesize 0.78} & {\footnotesize 0.69} & {\footnotesize 1.001} & {\footnotesize 1.855}\tabularnewline
{\footnotesize 30} & {\footnotesize 5} & {\footnotesize 150} & {\footnotesize 0.73} & {\footnotesize 0.73} & {\footnotesize 0.77} & {\footnotesize 0.69} & {\footnotesize 1.008} & {\footnotesize 1.938}\tabularnewline
{\footnotesize 40} & {\footnotesize 5} & {\footnotesize 200} & {\footnotesize 0.73} & {\footnotesize 0.73} & {\footnotesize 0.76} & {\footnotesize 0.7} & {\footnotesize 1.012} & {\footnotesize 1.987}\tabularnewline
{\footnotesize 50} & {\footnotesize 5} & {\footnotesize 250} & {\footnotesize 0.72} & {\footnotesize 0.72} & {\footnotesize 0.75} & {\footnotesize 0.69} & {\footnotesize 1.000} & {\footnotesize 1.993}\tabularnewline
{\footnotesize 100} & {\footnotesize 5} & {\footnotesize 500} & {\footnotesize 0.72} & {\footnotesize 0.72} & {\footnotesize 0.74} & {\footnotesize 0.7} & {\footnotesize 1.004} & {\footnotesize 2.076}\tabularnewline
{\footnotesize 5} & {\footnotesize 10} & {\footnotesize 50} & {\footnotesize 0.87} & {\footnotesize 0.88} & {\footnotesize 0.93} & {\footnotesize 0.83} & {\footnotesize 0.994} & {\footnotesize 1.671}\tabularnewline
{\footnotesize 10} & {\footnotesize 10} & {\footnotesize 100} & {\footnotesize 0.85} & {\footnotesize 0.86} & {\footnotesize 0.89} & {\footnotesize 0.82} & {\footnotesize 0.998} & {\footnotesize 1.941}\tabularnewline
{\footnotesize 20} & {\footnotesize 10} & {\footnotesize 200} & {\footnotesize 0.85} & {\footnotesize 0.85} & {\footnotesize 0.88} & {\footnotesize 0.82} & {\footnotesize 1.009} & {\footnotesize 2.160}\tabularnewline
{\footnotesize 30} & {\footnotesize 10} & {\footnotesize 300} & {\footnotesize 0.84} & {\footnotesize 0.84} & {\footnotesize 0.87} & {\footnotesize 0.82} & {\footnotesize 1.001} & {\footnotesize 2.230}\tabularnewline
{\footnotesize 40} & {\footnotesize 10} & {\footnotesize 400} & {\footnotesize 0.84} & {\footnotesize 0.84} & {\footnotesize 0.86} & {\footnotesize 0.82} & {\footnotesize 1.003} & {\footnotesize 2.287}\tabularnewline
{\footnotesize 50} & {\footnotesize 10} & {\footnotesize 500} & {\footnotesize 0.84} & {\footnotesize 0.84} & {\footnotesize 0.86} & {\footnotesize 0.82} & {\footnotesize 1.004} & {\footnotesize 2.325}\tabularnewline
{\footnotesize 100} & {\footnotesize 10} & {\footnotesize 1000} & {\footnotesize 0.84} & {\footnotesize 0.84} & {\footnotesize 0.85} & {\footnotesize 0.82} & {\footnotesize 1.006} & {\footnotesize 2.422}\tabularnewline
{\footnotesize 5} & {\footnotesize 25} & {\footnotesize 125} & {\footnotesize 0.94} & {\footnotesize 0.95} & {\footnotesize 0.97} & {\footnotesize 0.92} & {\footnotesize 0.995} & {\footnotesize 1.806}\tabularnewline
{\footnotesize 10} & {\footnotesize 25} & {\footnotesize 250} & {\footnotesize 0.94} & {\footnotesize 0.94} & {\footnotesize 0.96} & {\footnotesize 0.92} & {\footnotesize 1.006} & {\footnotesize 2.147}\tabularnewline
{\footnotesize 20} & {\footnotesize 25} & {\footnotesize 500} & {\footnotesize 0.93} & {\footnotesize 0.93} & {\footnotesize 0.95} & {\footnotesize 0.92} & {\footnotesize 1.000} & {\footnotesize 2.363}\tabularnewline
{\footnotesize 30} & {\footnotesize 25} & {\footnotesize 750} & {\footnotesize 0.93} & {\footnotesize 0.93} & {\footnotesize 0.94} & {\footnotesize 0.92} & {\footnotesize 1.002} & {\footnotesize 2.468}\tabularnewline
{\footnotesize 40} & {\footnotesize 25} & {\footnotesize 1000} & {\footnotesize 0.93} & {\footnotesize 0.93} & {\footnotesize 0.94} & {\footnotesize 0.92} & {\footnotesize 1.002} & {\footnotesize 2.531}\tabularnewline
{\footnotesize 50} & {\footnotesize 25} & {\footnotesize 1250} & {\footnotesize 0.93} & {\footnotesize 0.93} & {\footnotesize 0.94} & {\footnotesize 0.92} & {\footnotesize 1.003} & {\footnotesize 2.574}\tabularnewline
{\footnotesize 100} & {\footnotesize 25} & {\footnotesize 2500} & {\footnotesize 0.93} & {\footnotesize 0.93} & {\footnotesize 0.93} & {\footnotesize 0.92} & {\footnotesize 1.004} & {\footnotesize 2.681}\tabularnewline
\hline 
\end{tabular}\caption{The Johnson \& Rust (1992) simulation results for Satterthwaite's
(1946) effective degrees of freedom with improved approximation and
current NAEP adjustment.}
\foreignlanguage{american}{\label{j_and_r_table1-1}}\selectlanguage{american}%
\end{table}

\section*{References}

American Institutes of Research (n.d.) Analyzing NCES Data Using EdSurvey:
A User’s Guide. Section 11.4. Estimation of the Degrees of Freedom.
https://naep-research.airprojects.org/portals/0/edsurvey\_a\_users\_guide/\_book/
methods.html?q=degrees\%20of\%20freedom\#estimation-of-degrees-of-freedom

Johnson, E. G., \& Rust, K. F. (1992). Population Inferences and Variance
Estimation for NAEP Data. Journal of Educational Statistics, 17(2),
175--190. doi:10.2307/1165168

Johnson, E.G., and Rust, K.F. (1993). Effective Degrees of Freedom
for Variance Estimates From a Complex Sample Survey. American Statistical
Association 1993 Proceedings: Survey Research Methods Section, pp.
863--866.

Johnson, E. \& Rust, K. (1992). Effective Degrees of Freedom for Variance
Estimates from a Complex Sample Survey, Proceedings of the Section
on Survey Research Methods, American Statistical Association. (preprint
copy obtained from K. Rust, September, 2024).

Lipsitz, S., Parzen, M., \& Zhao, L. P. (2002). A Degrees-Of-Freedom
approximation in Multiple imputation. Journal of Statistical Computation
and Simulation, 72(4), 309--318. https://doi.org/10.1080/00949650212848

NCES (n.d.) NAEP Technical Documentation: Estimation of the Degrees
of Freedom. https://nces.ed.gov/ \\
nationsreportcard/tdw/analysis/2000\_2001/infer\_ttest\_df.aspx 

Nedelman, J. R. \& Jia, X. (1998) An extension of Satterthwaite's
approximation applied to pharmacokinetics. Journal of Biopharmaceutical
Statistics, 8:2, 317-328, doi:10.1080/10543409808835241

Rubin, D. B., \& Schenker, N. (1986). Multiple Imputation for Interval
Estimation From Simple Random Samples With Ignorable Nonresponse.
Journal of the American Statistical Association, 81(394), 366--374.
https://doi.org/10.2307/2289225

Satterthwaite, F. E. (1941). Synthesis of variance. Psychometrika,
6(5), 309-316. doi:10.1007/BF02288586

Satterthwaite, F. E. (1946). An Approximate Distribution of Estimates
of Variance Components. Biometrics Bulletin, 2(6), 110--114. doi:10.2307/3002019

Qian, J. (1998). Estimation of the effective degrees of freedom in
T-type tests for complex data. Proceedings of the Section on Survey
Research Methods, American Statistical Association, 704-708.

Welch, B. L. (1947). The generalization of `Student’s’ problem when
several different population variances are involved. Biometrika, 34(1/2),
28-35. https://doi.org/10.2307/2332510
\end{document}